\documentclass[prb,aps,fleqn,showpacs,reprint,twocolumn]{revtex4-1}

\usepackage{amssymb}
\usepackage{amsmath}
\usepackage{graphicx}
\usepackage{bm}

\begin{document}

\title{Highly sensitive gas and temperature sensor based on conductance modulation in graphene with multiple magnetic barriers}

\author{Nojoon Myoung}
\email{nmyoung@cc.uoi.gr} \affiliation{Department of Material
Science and Engineering, University of Ioannina, Ioannina 45110,
Greece}
\author{Elefterios Lidorikis}
\affiliation{Department of Material Science and Engineering,
University of Ioannina, Ioannina 45110, Greece}

\date{\today}

\begin{abstract}
The electronic and transport properties of graphene modulated by
magnetic barrier arrays are derived for finite temperature.
Prominent conductance gaps, originating from quantum interference
effects are found in the periodic array case. When a structural
defect is inserted in the array, sharp defect modes of high
conductance appear within the conductance gaps. These modes can be
shifted by local doping in the defect region resulting into sensing of
the chemical molecules that adhere on the graphene sheet.
 In general it is found that sensitivity is strongly
dependent on temperature due to smoothing
out of the defect-induced peaks and transport gaps. This
temperature dependence, however, offers the added capability for
sub-mK temperature sensing resolution, and thus an opportunity towards
ultra-sensitive combined electrochemical-calorimetric sensing.
\end{abstract}


\maketitle

\section{Introduction}

Advances in fabrication techniques since the isolation of monolayer
graphene\cite{Novoselov2004} have allowed the realization of a
variety of graphene-based applications\cite{Novoselov2012, Ferrari2015Nanoscale}.
In terms of electronic applications, graphene's high carrier
mobility\cite{Bolotin2008,Mayorov2011} has enabled its application
as an electrical conducting
channel\cite{Li2008,Geim2009,Schwierz2010}, however, the lack of
strong conductance modulation\cite{Williams2007} due to Klein
tunneling\cite{Klein1929,Cheianov2006,Katsnelson2006} has limited
its utilization for other graphene-based electronic devices. The
modulation of charge carriers in graphene is still a significant
issue of ongoing research for graphene
nanoelectronics\cite{Moriya2014,Shih2014}.

A conductance modulation in graphene can be introduced by utilizing
various mechanisms that open an electronic band gap, such as the
interaction with
susbtrates\cite{Giovannetti2007,Usachov2010,Decker2011,Bokdam2011},
elastic strain\cite{Ni2008,Shemella2009,Kim2009}, or finite-size
effects in graphene nanoribbons\cite{Han2007,Baringhaus2014}. Aside
from the band gap opening, however, these methods also introduce
unavoidable disorders, such as charge impurities or structural
imperfections by
substrates\cite{Ando2006,Fratini2008,Chen2008,Katsnelson2007} and
strong backscattering by rough edges of graphene
nanoribbons\cite{Fang2008,Wang2008,Mocciolo2009}, that deteriorate
the transport properties of graphene. An alternative mechanism for
band-gap opening, that in principle does not produce any disorder
effects in graphene, is the use of inhomogeneous magnetic fields.
Particularly, since inhomogeneous magnetic fields can tune the Dirac
fermion transport in graphene mimicking a typical potential barrier
for Dirac fermions\cite{RamenzaniMasir2008,Ghosh2009}, one can
produce magnetic confinement effects for graphene
carriers\cite{DeMartino2007,Myoung2011a}. These effects have
motivated research on various structures, e.g., magnetically defined
quantum dots\cite{DeMartino2007,Schnez2008}, Dirac fermion
waveguides\cite{Myoung2011a,Ghosh2008}, and
superlattices\cite{Wu2008,DellAnna2009}.

Besides nanoelectronics, another important aspect of graphene
applications is gas sensing\cite{Barbolina2006,Schedin2007}. The
operation principle of a graphene gas sensor relies on measuring the
modulated transport properties of graphene that are induced by the
adhesion of gas molecules on the graphene surface. In particular,
conductance changes arise from the injection of charged carriers
into graphene from adhered gas molecules such as O$_2$, NO$_2$ or
NH$_3$, which play roles of electron donors or
acceptors\cite{Ohta2006,Schedin2007,Zhou2008}. It has been shown
that highly sensitive graphene-based gas sensing capable of
detecting individual molecule adhesion\cite{Schedin2007} is
possible, but a large array of sensors might still be required in
order to enlarge the exposed area of graphene and thus increase  the
chances for molecule adhesion at shot-time exposures and minute
concentrations. Alternatively, here we explore ways of increasing
the minimum exposed area required for single molecule detection
without making arrays, which is significant for minimizing device
sizes and detection speeds.

\begin{figure*}[hbtp!]
\includegraphics[width=17cm]{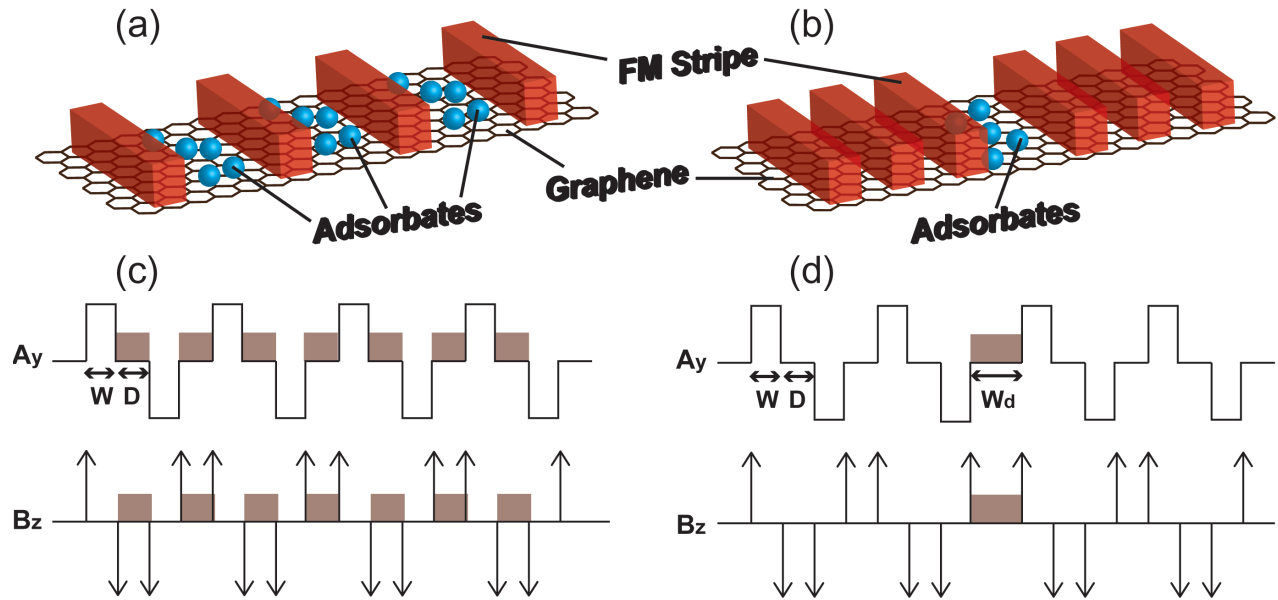}
\caption{ (a) and (b) Schematic views of magnetic barriers
considered in the zero-temperature limit and finite-temperature
regime. A series of ferromagnetic (FM) stripes (red blocks) are
placed upon graphene sheet, and adsorbates (blue spheres) can be
adhered in specific regions. (c) and (d) Magnetic vector potentials
$A_{y}$ and their corresponding magnetic field $B_{z}$ profiles for
the magnetic structures (a) and (b), respectively. Shaded boxes
implies adsorbate-induced local doping potentials. The magnetic
structures are characterized by the barrier width $W$, the
inter-barrier distance $D$, and the width of the defect region
$W_{d}$.}\label{fg:model}
\end{figure*}

Here, we theoretically study an alternative possibility for
achieving highly sensitive gas sensing, which is to utilize the
adsorbate-induced electrochemical doping effects in order to
modulate the tunneling resonances (TR) and transport band gaps (TBG)
emerging in magnetic barrier arrays in graphene. We show that in
such graphene sensors the presence of adsorbates introduces strong
 changes into graphene's electrical conductance. We take
into account two structures for the conductance modulation, as
illustrated by Fig.~\ref{fg:model}. First, we consider a periodic
array of magnetic barriers in which the inter-barrier regions are
exposed to electrochemical adsorbates. In the second case, we
consider a periodic magnetic barrier array with a structural
defect, where electrochemical doping is induced only in the defect
region. We investigate the ballistic conductance and its modulation
by doping effects for both structures, in different temperature
ranges. Specifically, in the zero-temperature limit, the presence of
a structural defect leads to sharp resonance peaks in the
conductance spectra, that can be sensitively modulated by local
doping in the defect region. In finite-temperature cases, on the
other hand, the overall sensitivity
is reduced because of thermal smoothing,
but still remains significant within
the transport band gaps when a periodic magnetic barrier array with
electrochemical doping in-between the barriers is considered.
By the same token, however, the strong temperature dependence which is regarded as a weak
point in terms of gas sensing, can be used for temperature
sensing. The highest sensing ability is expected at lower
temperatures, and yields a promising sensing platform for applications in a
low-temperature environment.

This manuscript is organized in the following manner: in
Sec.~\ref{sec:formalism}, we present the model Hamiltonian employed
to represent multiple magnetic barriers, and we explain the transfer
matrix formalism which is valid in the ballistic regime at finite
temperatures. In Sec.~\ref{sec:defect-effects}, we discuss the
features of the defect-induced tunneling resonances in the
transmission spectra. Next, in Sec.~\ref{sec:cond-mod} we show the
calculated results for the sensing effect by local doping in the
zero-temperature limit, and in Sec.~\ref{sec:cond-mod10K} we discuss
the large conductance modulation beyond the zero-temperature limit
and its temperature dependence. Finally, section~\ref{sec:summary}
contains the conclusions and a summary of our results.

\section{Model Hamiltonian and Formalism}
\label{sec:formalism}

We consider a graphene sheet with a periodic array of magnetic
barriers. Magnetic barriers have been experimentally realized by
ferromagnetic stripes on top of a graphene sheet\cite{Cerchez2007}.
To describe Dirac fermion ballistic transport through the system we
will use the transfer matrix formalism. Starting with the simplest
case of a single magnetic barrier along the $y$-direction, the Dirac
Hamiltonian reads\cite{Myoung2009}
\begin{align}
H=
v_{F}\vec{\sigma}\cdot\left[\vec{p}+e\vec{A}\left(x\right)\right]+U\left(x\right),
\end{align}
where $v_{F}\approx 10^6$~m/s is the Fermi velocity of Dirac fermions,
and $\vec{\sigma}=\left(\sigma_{1},\sigma_{2}\right)$ are Pauli
matrices acting on sublattices of graphene. For simplicity, the magnetic barrier is
characterized by a rectangular profile of the vector potential:
\begin{align}
\vec{A}\left(x\right)=Bl_{B}\left[\Theta\left(x\right)-\Theta\left(x-W\right)\right]\hat{y},
\end{align}
where $B$ is the magnetic field strength,
$l_{B}=\sqrt{\hbar/eB}$ the characteristic magnetic length,
$\Theta\left(x\right)$ is the Heaviside step function and $W$ is
the width of the magnetic barrier. The corresponding magnetic fields
are given by $\vec{B}=\nabla\times\vec{A}$, which corresponds to
delta-function-like spikes of opposite sign at the two edges of the
magnetic barrier. A general expression of the solution in the three
different regions $x<0$, $0<x<W$ and $W<x$ is written as
\cite{Myoung2009}:
\begin{align}
\psi_{j}\left(x\right)=a_{j}e^{ik_{j}x}\left(\begin{array}{c}1\\s_{j}e^{i\phi_{j}}\end{array}\right)+b_{j}e^{-ik_{j}x}\left(\begin{array}{c}1\\-s_{j}e^{-i\phi_{j}}\end{array}\right),
\end{align}
where $j=1,2,$ or $3$ represents different regions. The solution can
be written in the matrix form:
\begin{align}
\psi_{j}\left(x\right)=\mathbf{Q}_{j}\left(x\right)\left(\begin{array}{c}a_{j}\\b_{j}\end{array}\right)
\end{align}
with the matrices $\mathbf{Q}_1=\mathbf{Q}_3 \equiv \mathbf{Q}$ and $\mathbf{Q}_2 \equiv \mathbf{Q}'$  defined  as:
\begin{align}
\mathbf{Q}\left(x\right)&= \left(\begin{array}{cc}e^{ik_{x}x}&e^{-ik_{x}x}\\se^{ik_{x}x}e^{i\phi}&-se^{-ik_{x}x}e^{-i\phi}\end{array}\right),\nonumber\\
\mathbf{Q}'\left(x\right)&=\left(\begin{array}{cc}e^{ik'_{x}x}&-e^{-ik'_{x}x}\\s'e^{ik'_{x}x}e^{i\phi '}&-s'e^{-ik'_{x}x}e^{-i\phi '}\end{array}\right),
\end{align}
where
\begin{align}
k_{x}&=\sqrt{\left(\frac{\epsilon-U}{\hbar v_{F}}\right)^{2}-k_{y}^{2}},\nonumber\\
k'_{x}&=\sqrt{\left(\frac{\epsilon}{\hbar v_{F}}\right)^{2}-\left(k_{y}+\frac{eA_y}{\hbar}\right)^{2}},\nonumber\\
\phi &=\tan^{-1}{\left(\frac{k_{y}}{k_{x}}\right)}, \qquad \phi '
=\tan^{-1}{\left(\frac{k_{y}+eA_y/\hbar}{k'_{x}}\right)},\nonumber\\
s&=\text{sgn}\left(\epsilon-U\right), \qquad
s'=\text{sgn}\left(\epsilon
\right).\label{eq:wavevector_definitions}
\end{align}

We take into account the low-lying
excitation of Dirac fermions in graphene, below $100$~meV from the
charge neutral point, in order to remain within the
linear-band approximation governed by the Dirac equation.
The magnetic barrier height is found from the definition of $k'_x$
at normal incidence to be $E_b=v_{F}eA_y$.
 To avoid having negligible tunneling of Dirac fermions through the magnetic barrier,
the magnetic barrier width should not exceed a few tens nm.
Additionally, the total lateral size of a multiple magnetic barrier array should not be
over a few microns so we could still be in ballistic transport regime
even at room-temperatures\cite{Mayorov2008}.

In order to obtain the transmission and reflection probabilities
through a single magnetic barrier, we calculate the undetermined
coefficients $a_{i}$ and $b_{i}$ through the application of the boundary
conditions, i.e. that wavefunctions should be continuous at the
interfaces $x=0$ and $x=W$.
We note that in this study we do not need to take into account the Zeeman energy
$\vec{\sigma}\cdot g\mu_{B}\vec{B}$, where $g$ is the gyromagnetic
factor for Dirac fermions in graphene and $\mu_{B}$ is the Bohr
magneton, because the anti-symmetric magnetic field profiles of each
single magnetic barrier yield no spin-dependent transport
phenomena\cite{Myoung2011b}. The wavefunction continuity provides the following equations:
\begin{align}
\mathbf{Q}\left(0\right)\left(\begin{array}{c}a_{1}\\b_{1}\end{array}\right)&=\mathbf{Q}'\left(0\right)\left(\begin{array}{c}a_{2}\\b_{2}\end{array}\right),\nonumber\\
\mathbf{Q}'\left(W\right)\left(\begin{array}{c}a_{2}\\b_{2}\end{array}\right)&=\mathbf{Q}\left(W\right)\left(\begin{array}{c}a_{3}\\b_{3}\end{array}\right).
\end{align}
By combining these, we get an equation connecting
the coefficients the incoming and outgoing solutions:
\begin{align}
\left(\begin{array}{c}a_{3}\\b_{3}\end{array}\right)=\mathbf{T}\left(\begin{array}{c}a_{1}\\b_{1}\end{array}\right),
\end{align}
where $\mathbf{T}$ is the transfer
matrix\cite{McKeller2007,Barbier2008}.
\begin{align}
\mathbf{T}&=\mathbf{Q}^{-1}\left(W\right)\mathbf{Q}'\left(W\right)\mathbf{Q}'^{-1}\left(0\right)\mathbf{Q}\left(0\right),\nonumber\\
&=\left(\begin{array}{cc}\left(t^{-1}\right)^{\ast}&rt^{-1}\\(rt^{-1})^*&t^{-1}\end{array}\right).
\end{align}
Alternatively, the transfer matrix could also be derived using
optical analogies, as shown in Appendix~\ref{app:trans_mat}. We note here that
since we are interested in ballistic
transport of Dirac fermions through the structures, no disorder effects
that could lead to energy dissipation by inelastic scattering were included
in the our formalism.

The transmission probability $\mathcal{T}$ is obtained from the transfer matrix by
$\mathcal{T}=\left|t\right|^{2}$. Note that $\mathcal{T}$ is not only a function of the energy
but also of the incident angle of the incoming Dirac
fermions. In the ballistic regime and at finite temperature, the
conductance through a two-dimensional system is obtained by the
weighted average of the transmission function over the incident
angle, in accordance to the Landauer-B\"{u}ttiker
formalism\cite{Buttiker1985}:
\begin{align}
G\left(E_{F}\right)&=\frac{4e^{2}L_{y}}{\pi^{2}\hbar^{2}v_{F}}\int_{-\infty}^{+\infty}\int_{-\pi/2}^{+\pi/2}\epsilon\mathcal{T}\left(\epsilon,\phi\right) \cos{\phi}\nonumber\\
&\times\left(-\frac{\partial f}{\partial \epsilon}\right)d\phi
d\epsilon,\label{eq:cond_finiteK}
\end{align}
where $L_{y}$ is the system size in the transverse ($y$) direction,
$\phi$ is the propagating direction of incoming Dirac fermions, and
$f\left(\epsilon,E_{F},T\right)=\left\{1+\exp\left[{\left(\epsilon-E_{F}\right)/k_{B}T}\right]\right\}^{-1}$
is Fermi-Dirac distribution with given Fermi energy $E_{F}$ and
temperature $T$. In the zero temperature limit
 the conductance formula is simplified to:
\begin{align}
G\left(E_{F}\right)=\frac{4e^{2}L_{y}E_{F}}{\pi^{2}\hbar^{2}v_{F}}\int_{-\pi/2}^{+\pi/2}\mathcal{T}\left(E_{F},\phi\right)\cos{\phi}d\phi.\label{eq:cond_0K}
\end{align}
Next, we consider a monolayer graphene sheet in the presence of a
series of magnetic barriers. The models considered in this study are
shown in Fig.~\ref{fg:model}, where we assume periodically
arranged barriers with an inter-barrier distance $D$. Note that we
set an alternating configuration of the magnetic barriers, i.e.,
alternating signs of the vector potentials. The transfer matrix
formalism for multiple magnetic barriers can be extended from the
case of the single magnetic barrier:
\begin{align}
\mathcal{\mathbf{T}}=\prod_{i=0}^{N}\mathbf{T}_{i},
\end{align}
where $N$ is the number of magnetic barriers in use, and
\begin{align}
\mathbf{T}_{i}=\left\{\begin{array}{ll}\mathbf{Q}^{-1}\left(x_{i}+W\right)\mathbf{Q}'_{+}\left(x_{i}+W\right)&\\
\qquad\times\mathbf{Q}_{+}^{\prime -1}\left(x_{i}\right)\mathbf{Q}\left(x_{i}\right),\qquad
i~\mbox{is
odd}\\
\mathbf{Q}^{-1}\left(x_{i}+W\right)\mathbf{Q}'_{-}\left(x_{i}+W\right)&\\
\qquad\times\mathbf{Q}_{-}^{\prime -1}\left(x_{i}\right)\mathbf{Q}\left(x_{i}\right),\qquad
i~\mbox{is even}\end{array}\right..
\end{align}
where $x_{i}=i\left(W+D\right)$ is the position of $i$-th
magnetic barrier and $\mathbf{Q}'_{\pm}$ is for upward
or downward magnetic barriers. In the following sections we assume
20 magnetic barriers periodically arranged with $W=50$~nm,
$D=50$~nm, $E_b =v_{F}eA_{y}=20$~meV and
$W_{d}=100$~nm.

\section{Transmission probability through multiple magnetic barriers} \label{sec:defect-effects}

\begin{figure}[hbtp!]
\includegraphics[width=8.5cm]{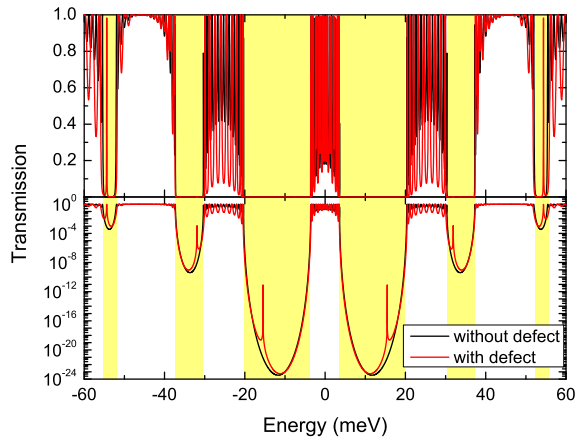}
\caption{ Transmission probabilities for the normal incidence versus
Fermi energy through multiple magnetic barriers with and without a
structural defect, in linear and logarithmic scales. Shaded regions
represent transporting gaps.}\label{fg:trans_defect}
\end{figure}

In this section, we discuss defect-induced resonances and
local-doping effects in the transmission probability of Dirac
fermions through multiple magnetic barriers.  The
existence of defect-induced resonances is best demonstrated by comparing the transmission
spectra through a periodic array of magnetic barriers without and with a structural defect, as illustrated in Fig.
\ref{fg:model}(a) and (b), respectively, and plotted for normal incidence in Fig.~\ref{fg:trans_defect}.

It is shown that the periodic arrangement of
magnetic barriers leads to the existence of transport gaps (TG) in
specific energy ranges, indicated by shaded regions in Fig.
\ref{fg:trans_defect}. The emergence of TGs comes from the combination of
 strong back scattering by the magnetic potentials and
quantum interference effects in the periodic structure, similarly to
the Kronig-Penny model. (The existence of the TG is best understood
by the band structure of the infinite magnetic barrier array, shown
in Appendix~\ref{app:band_structure}.) At energies below the magnetic
barrier height ($\epsilon<20$~meV), only resonant tunnelings through
the multiple magnetic barrier are allowed, with the number of
resonances being equal to the number of magnetic barriers. Let us
call these resonances for under-barrier tunneling `bound-state
tunneling resonances' (BTRs). For the defected magnetic barrier
array, on the other hand,
 sharp transmission peaks emerge in the TGs. We call these `defect-induced tunneling resonances' (DTRs).
Due to their positioning within the TGs, a large modulation in transmission probability is
expected near their peaks.


We next consider the effect of doping induced by adsorbates on the
graphene sheet. In the periodic array case we assume that all the
inter-barrier regions are exposed to adsorbates, while in the
defected array case we assume local doping applied to the defect
region only, as respectively illustrated in Fig. \ref{fg:model}(a)
and (b). Doping is introduced in our model as a rectangular
electrostatic potential barrier, characterized by the barrier height
$U_{d}$.

\begin{figure}[hbtp!]
\includegraphics[width=8.5cm]{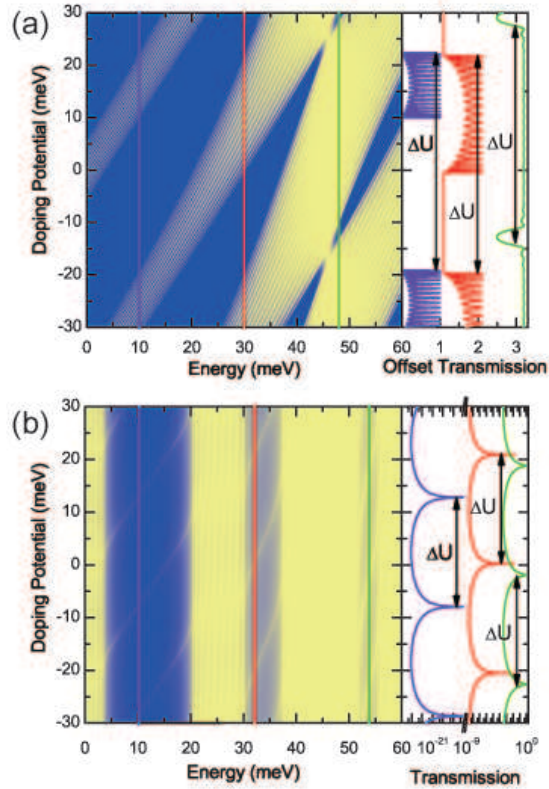}
\caption{ Effects of (a) the inter-barrier doping and (b) the local
doping in the defect region on transmission probability. 2D maps of
transmission are depicted as functions of Fermi energy and doping
potential in the case of normal incidence. The right insets display
the doping-dependent shift of the transmission peaks at given Fermi
energies, corresponding to the colored vertical lines.}
\label{fg:doping_trans}
\end{figure}

The effects of doping on the transmission
probability are shown in Fig.~\ref{fg:doping_trans}. In the case of inter-barrier doping of a periodic array
 without structural defects, the transmission spectra are entirely shifted by the doping potential.
 This shift is enough to produce strong qualitative changes in the transmission spectra
 above the barrier height, e.g. at the specific energy
$\epsilon\sim46$~meV, where the TG disappears and the transmission of Dirac
fermions is almost unaffected by doping. Local doping in the defect region,
 on the other hand, does not lead to a
global shift of the transmission spectra, but produces a clear shift
of the DTRs: the transmission peaks are shifted, and new ones
periodically appear as the local doping potential increases.

The periodic nature of the transmission peaks is
well interpreted by the quantum phase through the regions
where doping potentials are induced. The round-trip phase acquired by Dirac
fermions while moving through a distance $d$ is given by
\begin{align}
\varphi=2d\sqrt{\left(\frac{\epsilon-U}{\hbar
v_{F}}\right)^{2}-k_{y}^{2}}.\label{eq:qphase}
\end{align}
At a given energy, the periodicity of the potential energy specific
strength $\Delta U$ that yields $\Delta\varphi=2\pi$ can be found.
In the particularly simple case of normal incidence,
Eq.~(\ref{eq:qphase}) leads to a universal $\Delta U=\pi\hbar
v_{F}/d$ where $d=D$ for the inter-barrier doping case and $W_{d}$
for the local doping case. Indeed, for the inter-barrier doping, we
find $\Delta U=41.4$~meV, as verified in the inset of
Fig.~\ref{fg:doping_trans}(a), while for the local doping the energy
interval between the defect-induced transmission peaks is $\Delta
U=20.7$~meV, also shown in inset of Fig. \ref{fg:doping_trans}(b).
The same insets show the universality of the doping potential
periodicity, being the same for different energies at normal
incidence, albeit it will vary according to the incident angle.

\begin{figure}[hbtp!]
\includegraphics[width=8.5cm]{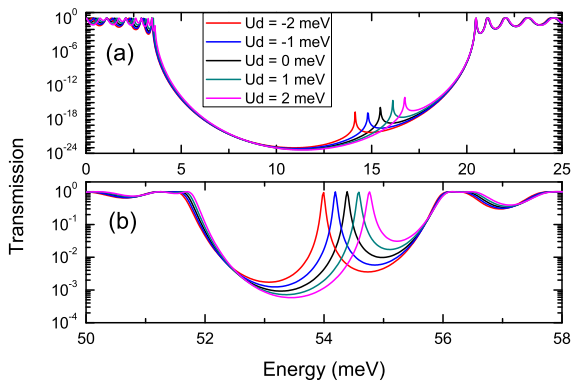}
\caption{ Local-doping dependence on transmission spectra around
different transporting gaps as a function of Fermi energy, in (a)
the under-barrier tunneling and (b) the over-barrier tunneling
regimes. The magnitude of the transmission probability is depicted
in logarithmic scale.}\label{fg:doping_dtr}
\end{figure}

The high sensitivity of the DTRs to small local doping is shown in
Fig.~\ref{fg:doping_dtr}, in two different energy ranges
corresponding to TGs in under- ($\epsilon<E_{b}$) and over-barrier
($\epsilon>E_{b}$) tunneling regimes. Owing to their positioning
within a TG, DTRs are sufficiently sharp, especially those within
the lower energy TG, so that one can expect that the transport
properties of graphene can be greatly modified by local doping. Note
that a doping strength of $U_{d}=1$~meV at $E_{F}=54$~meV
approximately corresponds to a change in carrier concentration
$\Delta n=8\times 10^{9}$~cm$^{-2}$, which is much smaller than the
actual carrier concentration of graphene $\sim
2\times10^{11}$~cm$^{-2}$ at that Fermi energy. (See Ref.
\cite{Hwang2007} and reference therein)

\section{Conductance Modulation by Doping at Zero Temperature}
\label{sec:cond-mod}

We studied in the previous section the transmission probabilities
of Dirac fermions through periodic magnetic barrier structures in
the presence or absence of structural defects, and found
that slight shifts of the defect-induced resonant peaks can
produce significant changes to the transmission probability.
In this section, we calculate the zero temperature ballistic conductance
through a graphene sheet decorated with magnetic barrier arrays
and its dependence on doping, with particular focus on the DTR shifts.

\begin{figure}[hbtp!]
\includegraphics[width=8.5cm]{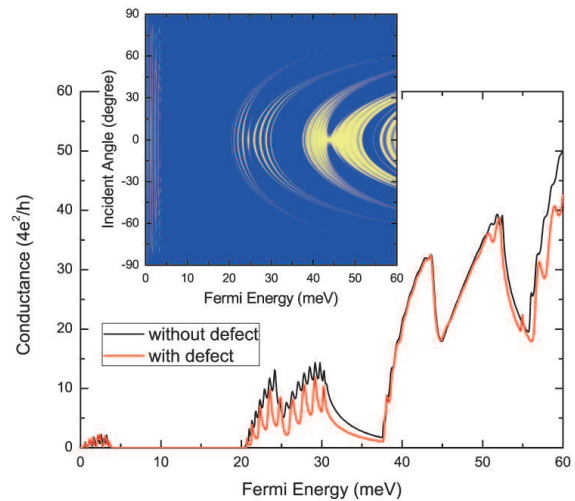}
\caption{ Ballistic conductance as a function of Fermi energy
through multiple magnetic structure with and without a structural
defect. $L_{y}=1$ $\mu m$ is taken into account for the calculation.
The inset exhibits the transmission function taken into account the
calculation of the conductance in the presence of the
defect.}\label{fg:cond_defect}
\end{figure}

In the zero-temperature limit, conductance is calculated from Eq.
(\ref{eq:cond_0K}), which we plot in Fig.~\ref{fg:cond_defect}.
Similarly to the transmission spectra, defect-induced peaks that are
modulated by local doping appear within the transport dips of the
conductance curve. There is, however, an important difference: the
conductance no longer drops to low values, especially at higher
energies. At low energy, on the other hand, the conductance is
dominated by under-barrier tunneling through the barriers, resulting
into a complete TG within which the conductance values almost
vanish. The complete gap in low energy is due to the strong
backscattering of the magnetic barriers, while the incomplete dips
in high energies originate from the quantum interference effects in
the periodic array of magnetic barriers. The nature of the two
transport regimes is further revealed by the symmetry of the
conductance resonances as shown in the inset in Fig.
\ref{fg:cond_defect} in connection to Eq.~(\ref{eq:cond_0K}): the
isotropic backscattering from the alternating magnetic barriers in
the low-energy under-barrier tunneling regime reflects a complete
transport gap, while the strong dependence on the incident angle of
the high-energy over-barrier tunneling of Dirac fermions reflects
the incomplete drop of conductance values.

With the possibility of a sensing application, it is necessary to
have a figure of merit for the conductance sensitivity to doping. To
this end, we introduce gauge factors ($\mathcal{F}$) as commonly
used for characterizing sensing performance:
\begin{align}
\mathcal{F}_{R}=\frac{\Delta R/R}{\Delta
n/n},\qquad\mathcal{F}_{G}=\frac{\Delta G/G}{\Delta
n/n},\label{eq:gauge_factor}
\end{align}
where $R=1/G$ is the graphene resistance and
$n={\left(E_F+U_{d}\right)^{2}}/(\pi\hbar^{2}v_{F}^{2})$ the local
carrier concentration as a consequence of the local doping. These
two different definitions of the gauge factor can be selectively
applied to specify the sensitivity, according to the way of
measuring the transport properties of graphene.

\subsection{Effects of doping on the ballistic
conductance without a defect}

\begin{figure}[hbtp!]
\includegraphics[width=8.5cm]{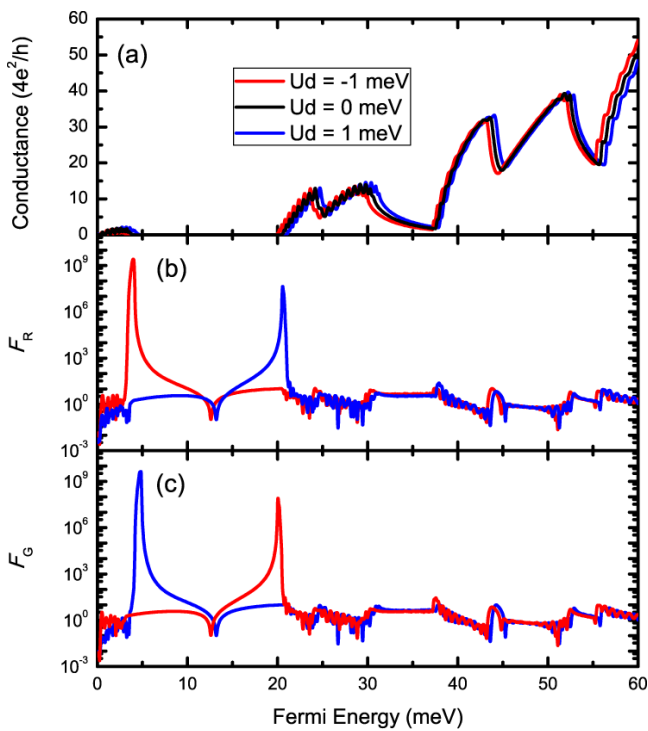}
\caption{ (a) Ballistic conductance as a function of Fermi energy
through periodic magnetic barriers, for different doping potentials.
(b) and (c) Corresponding gauge factors $\mathcal{F}_{R}$ and
$\mathcal{F}_{G}$ as a function of Fermi energy for different doping
potentials.}\label{fg:cond_PMBs}
\end{figure}

For a periodic array of magnetic barriers, the ballistic conductance
calculated by Eq. (\ref{eq:cond_0K}) and its dependence on the
inter-barrier doping are displayed in Fig. \ref{fg:cond_PMBs}(a).
The conductance curves are entirely shifted as the doping potential
varies. Figures~\ref{fg:cond_PMBs}(b) and (c) present the gauge
factors $\mathcal{F}_{R}$ and $\mathcal{F}_{G}$, i.e. the
sensitivity of the conductance modulation by the inter-barrier
doping, for various doping potentials. As expected, the
$\mathcal{F}_{G}$ exhibits an opposite dependence on the doping
potential compared to $\mathcal{F}_{R}$, according to their
definition (see Eq.~(\ref{eq:gauge_factor})). Since the
inter-barrier doping effects lead to the entire shift of the
conductance, very sensitive changes in the conductance value are
obtained around the BTRs and the complete TG edges in the
under-barrier tunneling regime. A large gauge factor $\sim10^{8}$
is obtained near the TG edge $E_{F}\sim20$ meV, resulting into
$\Delta n=\left(n/\mathcal{F}_{R}\right)\left(\Delta
R/R\right)\cong30~\mbox{cm}^{-2}$ for $\Delta R/R=0.1$, i.e. assuming
 a $10~\%$ resistance variation as the measurement resolution.
Even higher gauge factor is obtained in the low energy edge at $E_{F}\sim5$ meV.  In other words, such a large
value of the gauge factor means that the detection of the single
free carrier injection/extraction is possible within
$3.3~\mbox{mm}^{2}$ area exposed to electrochemical adhesion for
$E_{F}=20~\mbox{mev}$.  In
the over-barrier regime, on the other hand, the sensitivity of the
conductance modulation by the inter-barrier doping is expected to be
much less sensitive, compared to the under-barrier regime.

\subsection{Sensitivity of the conductance modulation with a defect: under-barrier versus over-barrier}
\label{sec:cond-mod0K}

\begin{figure}[hbtp!]
\includegraphics[width=8.5cm]{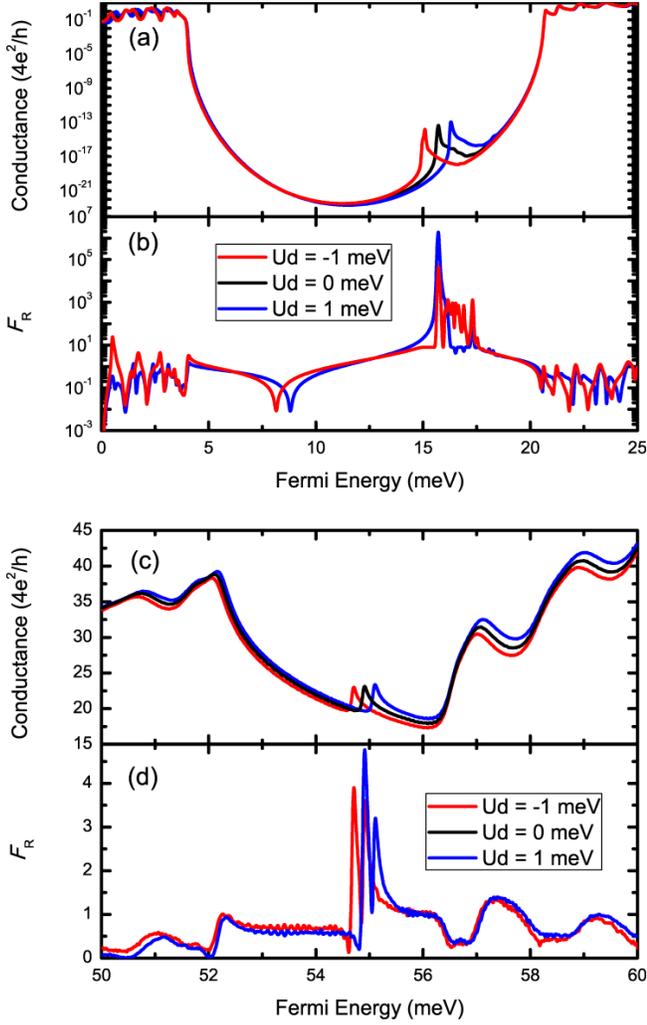}
\caption{ (a) Local-doping dependence of the ballistic conductance
as a function of Fermi energy through multiple magnetic barriers
with the structural defect and (b) its corresponding gauge factor
$\mathcal{F}_{R}$, for under-barrier tunneling. (c) and (d) The same
plots as (a) and (b) but for over-barrier
tunneling.}\label{fg:gauge_factor}
\end{figure}

Next, we discuss the effects of  local doping on the DTRs. In
Figs.~\ref{fg:gauge_factor}(a) and (c), the local-doping dependence
of the conductance is shown in the under-barrier and over-barrier
tunneling regimes, respectively, exhibiting shifts similarly to the
transmission peaks at normal incidence. Despite the small change in
the doping potential ($U_{d}=1$~meV), the separation of the shifted
peaks are enough to produce a significant amount of conductance
modulation, offering a scheme for sensitive adsorbate detection.

In the under-barrier tunneling regime, Fig.~\ref{fg:gauge_factor}(a)
exhibits a large contrast in the conductance values around DTRs. One
can expect that the conductance abruptly decreases with small amount
of doping because of the sharpness of the conductance peak in this
regime. This ultra-sensitive conductance modulation allows us to
obtain a very large value of the gauge factor, up to $\sim10^{6}$ as displayed in
Fig.~\ref{fg:gauge_factor}(b), but the actual magnitude of the conductance in this regime is
too small.

On the other hand, in the over-barrier tunneling regime, Fig.
\ref{fg:gauge_factor}(c) exhibits that the conductance values are
greater than those in the under-barrier regime and there are DTRs
within the conductance dip. The DTRs are shifted by the local doping
effects as well, but the gauge factor in the over-barrier regime is
expected to be much smaller, compared to those in the under-barrier
tunneling regime, as shown in Fig. \ref{fg:gauge_factor}(d),  due to
 the incomplete conductance drop, and thus small contrast,  around the DTRs.
 Here, we note that for simplicity we show $\mathcal{F}_{R}$ only, since
 maximum values of $\mathcal{F}_{G}$ are also found near the
DTRs and as expected are almost the same as those of $\mathcal{F}_{R}$.

\begin{figure}[hbtp!]
\includegraphics[width=8.5cm]{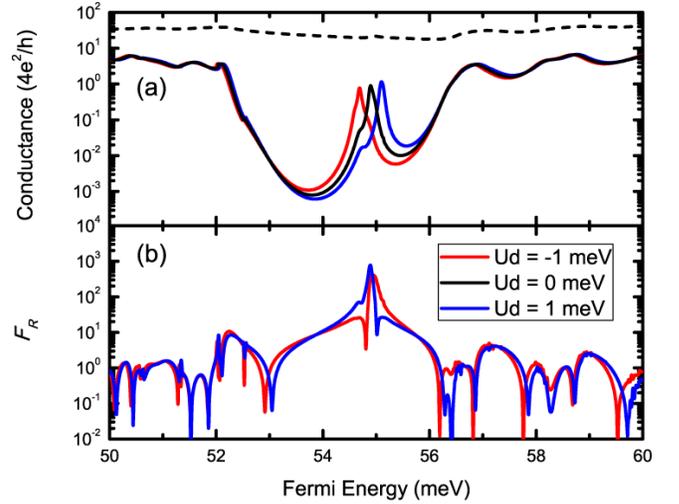}
\caption{ (a) Effects of the collimators on the conductance as a
function of Fermi energy. Dashed line implies the conductance curve
without the collimators, while solid lines indicate the conductance
curves for different local doping potentials in the defect region.
(b) Gauge factor $\mathcal{F}_{R}$ versus fermi energy,
corresponding to the doping potentials.}\label{fg:collimation}
\end{figure}

\subsection{Enhancement of the sensitivity by collimators}

The incomplete conductance drop in the over-barrier regime is
attributed to the transport contributions coming from all incident
angles of Dirac fermions. If these angular contributions could be
suppressed, the transport dips will get deeper, resulting into a
larger contrast near the DTRs. To this end, we propose the use of a
Dirac fermion collimator to kill the non-zero incident angle
contributions to the conductance.
 In particular, it has been known
even a single electrostatic barrier can produce a collimation
effect\cite{Park2008,Barbier2010}. This is easily seen in the first
of Eq.~(\ref{eq:wavevector_definitions}), where for $E=U_0$ any
non-zero value of $k_y$ results into an imaginary $k_x$, and thus
into suppressed transmission. The normal incidence transmission, on
the other hand, remains  always protected by Klein tunneling. In our
model, we thus introduce an electrostatic barrier of potential
height $U_{c}$ and barrier width $W_{c}=800$ nm at both sides of the
magnetic barrier array.  This  leads to the suppression of all
non-zero angular contributions to the conductance near $E_{0}\approx
U_{c}$, and thus to an increase of the conductance contrast around
it.

Figure~\ref{fg:collimation} shows the effects of a collimator with
potential height set to match the energy corresponding to a DTR.
Indeed, the transmission dip becomes deeper in the presence of the
collimator, increasing the conductance contrast, and thus
$\mathcal{F}_{R}$ becomes large up to $\sim10^{3}$. Such a
 gauge factor enables the detection of $\Delta
n=2.24\times10^{7}~\mbox{cm}^{-2}$, for $\Delta R/R=0.1$. As
experimentally demonstrated\cite{Schedin2007}, the exposure of a
graphene sheet to 1 $p.p.m.$ NO$_{2}$ gas results into $\Delta
n\sim5\times10^{10}$~cm$^{-2}$ with few $\%$ of resistance changes.
With the same order of magnitude of $\Delta R/R$, $\Delta
n=2.24\times10^{8}~\mbox{cm}^{2}$ is expected to be detected.
Therefore, our results offer an opportunity for sub-$p.p.m.$ level
detection of gas molecules by using the doping-induced shift of
DTRs. However, this is still much lower than the sensitivity offered
by the periodic array shown in the previous subsection, because  of
the still relatively large background conductance around the DTR,
despite the use of the collimator. In closing this section, we
remind that the application of the collimator to the under-barrier
regime would not be useful because complete TGs already exist there.

\section{Conductance Modulation by Doping Effects at Finite Temperature} \label{sec:cond-mod10K}

Our discussion regarding the conductance modulation has done in the
zero-temperature limit. It is necessary, however, to also know up to
what extend our results will be valid for the more practical cases
of finite temperature. In this section, we investigate
the finite-temperature effects on the ballistic transport through the two
types of multiple magnetic barrier structures and examine the conductance modulation
 by doping  based on Eq.~(\ref{eq:cond_finiteK}).

\subsection{Temperature effects on the conductance through periodic magnetic barriers}

We consider the periodic array of magnetic barriers with all inter-barrier regions exposed to electrochemical dopants (see Fig. \ref{fg:model}(a)).

\begin{figure}[hbtp!]
\includegraphics[width=8.5cm]{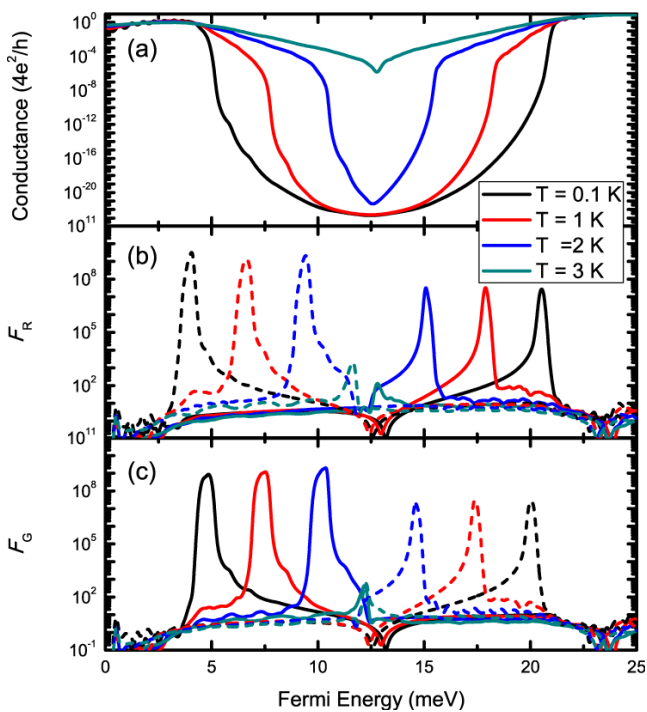}
\caption{ (a) Temperature dependence of the ballistic conductance
through periodic magnetic barriers as a function of fermi energy, in
the under-barrier tunneling regime. (b) and (c) Gauge factors
$\mathcal{F}_{R}$ and $\mathcal{F}_{G}$ corresponding to the
conductance spectra for different temperatures, with $U_{d}=1$ meV.
Dashed curves represent $\mathcal{F}_{R}$ and $\mathcal{F}_{G}$ with
$U_{d}=-1$ meV.}\label{fg:finite_gf_PMB}
\end{figure}

In the previous section, we found that the conductance
modulation for the periodic magnetic barriers is expected to be
significant around the edges of the complete TGs which correspond to
the under-barrier tunneling.  The calculated conductance modulation and the
corresponding sensitivity are depicted in Fig.~\ref{fg:finite_gf_PMB}, showing their temperature dependence. At
$T=0.1$ K, large values of the gauge factor are observed near the TG
edge, similarly with the results in the zero-temperature limit. This
gauge factor peaks are due to the fact that the conductance values
abruptly increase near the TG edges with a small $U_{d}$. As
temperature increases, the gauge factor peak is shifted to the
center of the TG, exhibiting minor reduction in the sensitivity up
to $2$ K. This shift of the gauge factor peak means that the
variation of the conductance values near the TG edge becomes
smoother as temperature increases. The sensitivity is dramatically
reduced at $T=3$ K, so that the sensitive detection of the presence
of the adsorbates in-between magnetic barriers is valid for the
low-temperature limit below $3$ K.

\subsection{Temperature effects on defect-induced tunneling
resonances}

As discussed, the key feature of  highly sensitive conductance
modulation in the defected array case results from the existence
of DTRs and  their shift by
doping. In order to consider the possibility for
finite-temperature gas sensing, it is necessary to see whether the DTRs
survive at finite temperatures.

\begin{figure}[hbtp!]
\includegraphics[width=8.5cm]{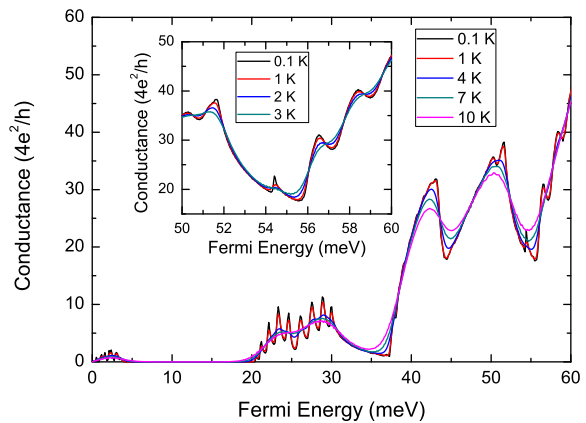}
\caption{ Conductance spectra as a function of Fermi energy at
different temperatures. Inset: detailed view of temperature
dependence of a defect-induced conductance
peak.}\label{fg:cond_finiteK}
\end{figure}

Figure~\ref{fg:cond_finiteK} displays the ballistic conductance
through multiple magnetic barriers with a defect as a function of
fermi energy for several temperatures. At low temperature $T=0.1$ K,
the distinct DTRs are clearly observed within the transporting dips,
similar to the zero-temperature limit. As temperature increases,
however, the DTRs, which are responsible for the
high sensitivity to the doping, become smoothed out. Thus, it is not expected to achieve
 high sensitivity with this system  at finite temperatures.

\begin{figure}[hbtp!]
\includegraphics[width=8.5cm]{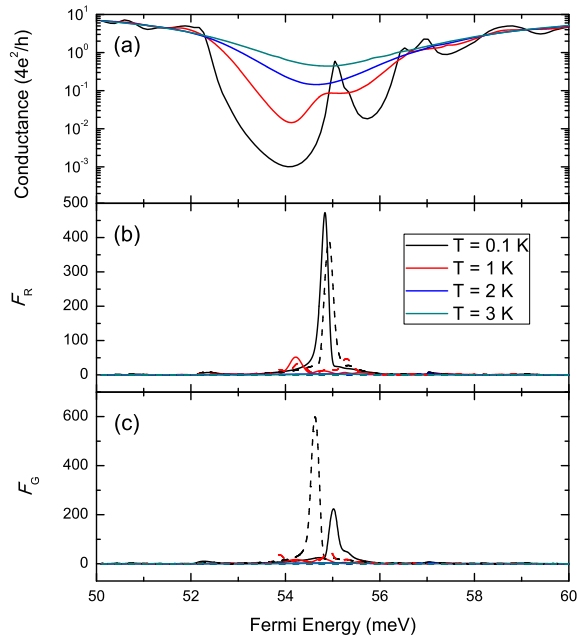}
\caption{ (a) Temperature-dependence of the collimator effects on
the ballistic conductance as a function of fermi energy around the
defect-induced peak in the over-barrier tunneling regime. (b) and
(c) Gauge factors $\mathcal{F}_{R}$ and $\mathcal{F}_{G}$
corresponding to the conductance spectra for different temperatures,
with $U_{d}=1$ meV. Dashed curves represent $\mathcal{F}_{R}$ and
$\mathcal{F}_{G}$ with $U_{d}=-1$ meV.}\label{fg:temp_gaugefactor}
\end{figure}

Indeed, the sensitivity of the conductance modulation near the DTR
severely diminishes as temperature increases,
as shown in Fig.~\ref{fg:temp_gaugefactor}(a).
 At low temperature $T=0.1$~K,
the doping dependence of the conductance still exhibits an apparent DTR
within the conductance drop, similarly to the zero-temperature limit. The
maximum value of $\mathcal{F}_{R}\sim500$ gives $\Delta
n=4.5\times10^{7}~\mbox{cm}^{-2}$, corresponding to a sensitivity reduction
of a factor of 2 compared to the zero-temperature case. Despite this reduction,
it is still enough
for sub-$p.p.m.$ level gas detection as aforementioned. However, as
temperature increases, the DTRs becomes completely smoothed out, eliminating
any sensitivity, as seen in Fig.~\ref{fg:temp_gaugefactor}(b) and (c).
Therefore, one can only expect highly sensitive detection of local doping at
the low-temperature limit below $1$ K. We note here, that the larger gauge factor
values are achieved for negative doping potentials because of the
asymmetric profile of the DTRs.

\subsection{Sensitive detection of temperature variation}

We have found profound deductions of the gauge factors at finite
temperatures/ However, while this is a limitation for gas sensing, such a severe temperature dependence may become an asset which allows us to detect minute temperature changes.

\begin{figure}[hbtp!]
\includegraphics[width=8.5cm]{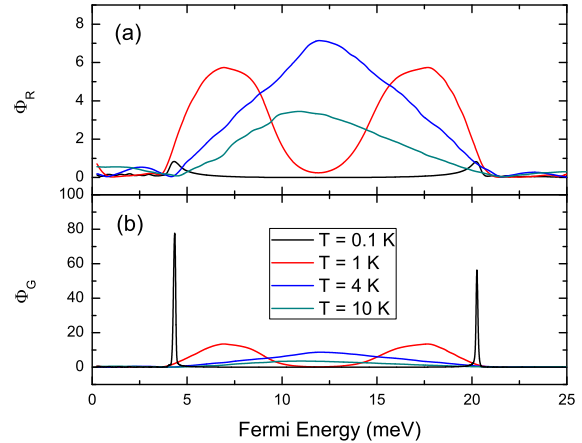}
\caption{ Gauge factors (a) $\Phi_{R}$ and (b) $\Phi_{G}$ as a
function of fermi energy with $\Delta T=0.1$ at different
temperatures.} \label{fg:gf_temp}
\end{figure}

In order to characterize the temperature sensing ability, we define
the following gauge factors with respect to temperature changes:
\begin{align}
\Phi_{R}=\frac{\Delta R/R}{\Delta T/T},\qquad\Phi_{G}=\frac{\Delta
G/G}{\Delta T/T},\label{eq:gf_temp}
\end{align}
where $\Delta T$ is the change in temperature. Similar to
$\mathcal{F}_{R}$ or $\mathcal{F}_{G}$, $\Phi_{R}$ and $\Phi_{G}$
imply how sensitively the conductance is modulated as temperature
changes. Figure \ref{fg:gf_temp} shows the calculated gauge factors
as a function of Fermi energy in the under-barrier tunneling regime.
It is clearly seen that $\Phi_{G}$ is suitable for the temperature
sensing. Large values of $\Phi_{G}$ are achieved near the TG edge
and the BTRs at low temperature $T=0.1$ K. As temperature increases,
the sharp and large gauge factor peaks are reduced and shifted to
the center of the TG because the TG becomes narrower and shallower
by temperature, as we aforementioned.

At specific temperatures, the detection of temperature change is
determined by $\Phi_{G}$ and $\Delta G/G$, as $\Delta
T=\left(T/\Phi_{G}\right)\left(\Delta G/G\right)$. For example, at
$T=0.1$ K, a $\Phi_{G}\sim80$ can be found, which for $10~\%$ conductance measurement
resolution results into sub-mK level
temperature detection. Obviously, the sensitivity of the temperature detection
becomes reduced at higher temperature, but the expected $\Delta T$
is still small below 1 K, assuming $\Delta G/G=0.1$. Therefore, the
strong temperature-dependent behavior of the ballistic conductance,
which reduces the potential for adsorbate detection, can be used for
temperature sensing effects.

\section{Summary}
\label{sec:summary}

This article assesses how the ballistic conductance through multiple
magnetic barriers is modulated by local doping induced via
electrochemical adsorbates on graphene surface. Both periodic and
defected barrier arrays were studied. In the zero-temperature limit,
large sensitivity to electrochemical doping with single molecule
detection was found at the edges of the transport gaps in the
periodic case under inter-barrier doping, while sub-\textit{p.p.m}
sensitivity was found on the defect-induced peaks of the defected
barrier array for local doping in the defect area only. In the
latter case, the use of a Dirac fermion beam collimator (i.e. a
suitable electric potential barrier) to suppress transport of Dirac
fermions with non-zero incident angles was found to be necessary.

The sensitivity of the conductance modulation is also discussed in the
finite-temperature case. The defect-induced transport resonances (DTR)
within the conductance gaps are
smoothed out as temperature increases, even around 1~K, and so that the sensitivity of
the conductance modulation becomes much less than that in
the zero-temperature limit. In the case of periodic barriers with inter-barrier
doping in the under-barrier tunneling regime,  large values of the sensitivity gauge factors
can still be found below 3 K, however, as temperature increases beyond 3 K, the gauge
factors drastically decreases because of the thermal smoothing.
 Interestingly, this strong temperature dependence of the
sensitivity is advantageous in terms of temperature sensing
effects, allowing  sub-K level temperature detection.
This temperature sensing capability is
found to increase as temperature decreases towards zero,
pointing  towards a  temperature sensor
that could be particularly useful in low-temperature experiments, outer space, etc. Furthermore,
the combination of the strong temperature dependence with the low thermal capacitance of the electron gas, may be an excellent platform for
extremely sensitive calorimetric studies involving energy transfer between adsorbates, optical transitions,
relaxation precesses, etc, in low-temperature experiments.

\begin{acknowledgements}
We acknowledge funding from EU Graphene Flagship (no. 604391).
\end{acknowledgements}


\appendix
\section{Optical analogy of the transfer matrix and analytic formalism}
\label{app:trans_mat}

For plane-wave solutions, the propagating behavior of Dirac fermions
can be governed by an optical analogy. The transfer matrix is also
expressed by an alternative representation.

As the starting point, let us consider a boundary $x=0$ between two
regions with different potentials. In the left and right sides of
the boundary, wavefunctions are given by
\begin{align}
\Psi_{0<x}\left(x\right)&=a_{1}e^{ik_{1}x}\left(\begin{array}{c}1\\s_{1}e^{i\phi_{1}}\end{array}\right)\nonumber\\
&+b_{1}e^{-ik_{1}x}\left(\begin{array}{c}1\\-s_{1}e^{-i\phi_{1}}\end{array}\right),\nonumber\\
\Psi_{0>x}\left(x\right)&=a_{2}e^{ik_{2}x}\left(\begin{array}{c}1\\s_{2}e^{i\phi_{2}}\end{array}\right)\nonumber\\
&+b_{2}e^{-ik_{2}x}\left(\begin{array}{c}1\\-s_{2}e^{-i\phi_{2}}\end{array}\right),
\end{align}
as well-interpreted in the main text. These plane-wave solutions
must be matched by the wavefunction continuity. The boundary
condition leads to the following equation:
\begin{align}
\left(\begin{array}{c}a_{2}\\b_{2}\end{array}\right)&=\frac{1}{2s_{2}\cos{\phi_{2}}}\left(\begin{array}{cc}s_{11}&s_{12}\\s_{21}&s_{22}\end{array}\right)\left(\begin{array}{c}a_{1}\\b_{1}\end{array}\right),\nonumber\\
&=\mathbf{I}_{1\leftarrow2}\left(\begin{array}{c}a_{1}\\b_{1}\end{array}\right),
\end{align}
where
\begin{align}
&s_{11}=s_{1}e^{i\phi_{1}}+s_{2}e^{-i\phi_{2}},
&&s_{12}=-s_{1}e^{-i\phi_{1}}+s_{2}e^{-i\phi_{2}},\nonumber\\
&s_{21}=-s_{1}e^{i\phi_{1}}+s_{2}e^{i\phi_{2}},
&&s_{22}=s_{1}e^{-i\phi_{1}}+s_{2}e^{i\phi_{2}}.
\end{align}
The `interface' matrix $\mathbf{I}_{1\leftarrow2}$ implies
scattering effects at the interface $x=0$ when Dirac fermions come
from the region 2 to the region 1. Next, we look at the propagation
of Dirac fermions through a region where potentials are
homogeneously given. The `propagation' matrix is given by
\begin{align}
\mathbf{P}_{i,l_{i}}=\left(\begin{array}{cc}e^{ik_{i}l_{i}}&0\\0&e^{-ik_{i}l_{i}}\end{array}\right),
\end{align}
where $l_{i}$ is the length of $i$-th region.

Now, we can study transmission of Dirac fermions through a potential
barrier by using the interface and the propagation matrix:
\begin{align}
\left(\begin{array}{c}a_{3}\\b_{3}\end{array}\right)&=\mathbf{P}_{3,l_{3}}\mathbf{I}_{3\leftarrow2}\mathbf{P}_{2,l_{2}}\mathbf{I}_{2\leftarrow1}\left(\begin{array}{c}a_{1}\\b_{1}\end{array}\right)\nonumber\\
&=\mathbf{M}\left(\begin{array}{c}a_{1}\\b_{1}\end{array}\right)
\end{align}
The matrix $\mathbf{M}$ corresponds to the transfer matrix that
describes the relation between incoming and outgoing waves. The
matrix elements are given by
\begin{align}
m_{11}&=e^{ik_{x}D}\left[\cos{\left(q_{x}W\right)}\right.\nonumber\\
&\left.+i\sin{\left(q_{x}W\right)}\left(\frac{ss'-\sin{\phi}\sin{\theta}}{\cos{\phi}\cos{\theta}}\right)\right],\nonumber\\
m_{12}&=e^{-ik_{x}D}\left[\cos{\left(q_{x}W\right)}\right.\nonumber\\
&\left.-i\sin{\left(q_{x}W\right)}\left(\frac{ss'-\sin{\phi}\sin{\theta}}{\cos{\phi}\cos{\theta}}\right)\right]=m_{11}^{\ast},\nonumber\\
m_{21}&=-e^{ik_{x}D-i\phi}\sin{\left(q_{x}W\right)}\left(\frac{ss'\sin{\phi}-\sin{\theta}}{\cos{\phi}\cos{\theta}}\right),\nonumber\\
m_{22}&=-e^{-ik_{x}D+i\phi}\sin{\left(q_{x}W\right)}\left(\frac{ss'\sin{\phi}-\sin{\theta}}{\cos{\phi}\cos{\theta}}\right),\nonumber\\
&=m_{21}^{\ast},
\end{align}
where all parameters $k_{x}$, $q_{x}$, $\phi$, $\theta$, $s_{1}$,
and $s_{2}$ are the same as those given in the main text. Also, $W$
and $D$ are the barrier and the inter-barrier width.

We next formulate the transmission problem:
\begin{align}
\left(\begin{array}{c}t\\0\end{array}\right)=\mathbf{M}\left(\begin{array}{c}1\\r\end{array}\right),
\end{align}
where $r$ and $t$ are the reflection and transmission coefficients.
Solving this equation, we obtain the following relations between $r$
and $t$:
\begin{align}
t=m_{11}+m_{12}r,\qquad 0=m_{21}+m_{22}r,
\end{align}
and we finally have:
\begin{align}
r=-\frac{m_{21}}{m_{22}},\qquad
t=\frac{\det\left[\mathbf{M}\right]}{m_{22}}.
\end{align}
In order to get $\det\left[\mathbf{M}\right]$, we take the
determinant of individual matrices as
$\det\left[\mathbf{A}\mathbf{B}\right]=\det\left[\mathbf{A}\right]\det\left[\mathbf{B}\right]$.
It is straightly seen $\det\left[\mathbf{P}\right]=1$, and the
determinant of the interfaces matrices is obtained by
\begin{align}
\det\left[\mathbf{I}_{2\leftarrow1}\right]=\frac{s\cos\phi}{s'\cos\theta},\qquad\det\left[\mathbf{I}_{1\leftarrow2}\right]=\frac{s'\cos\theta}{s\cos\phi}.
\end{align}
In results, the determinant of the transfer matrix is unity, as
expected for a lossless system.

Thus, the reflection and transmission probabilities are calculated
as follows:
\begin{widetext}
\begin{align}
R&=\left|r\right|^{2}=\frac{\sin\left(q_{x}W\right)\left(ss'\sin\phi-\sin\theta\right)^{2}}{\cos^{2}\phi\cos^{2}\theta\cos^{2}\left(q_{x}W\right)+\sin^{2}\left(q_{x}W\right)\left(ss'-\sin\phi\sin\theta\right)^{2}},\nonumber\\
T&=\left|t\right|^{2}=\frac{\cos^{2}\phi\cos^{2}\theta}{\cos^{2}\phi\cos^{2}\theta\cos^{2}\left(q_{x}W\right)+\sin^{2}\left(q_{x}W\right)\left(ss'-\sin\phi\sin\theta\right)^{2}},
\end{align}
\end{widetext}
It is easily found $R+T=1$, implying the flux conservation.

We formulate the transfer matrix which has the correspondence to the
photonic analogy of multilayer structures. We examine a periodic
array of magnetic barriers. In this case, Bloch's theorem leads to
the following expression:
\begin{align}
\left(\begin{array}{c}a_{3}\\b_{3}\end{array}\right)=\mathbf{M}\left(\begin{array}{c}a_{1}\\b_{1}\end{array}\right)=e^{\pm
iKL}\left(\begin{array}{c}a_{1}\\b_{1}\end{array}\right),
\end{align}
where $KL=k_{x}D+ss'q_{x}W$. We now have the secular equation:
\begin{align}
\left|\begin{array}{cc}m_{11}-e^{\pm
iKL}&m_{12}\\m_{21}&m_{22}-e^{\pm iKL}\end{array}\right|=0,
\end{align}
of which diagonalization gives the corresponding eigenstates:
\begin{align}
e^{\pm
iKL}=\frac{m_{11}+m_{22}\pm\sqrt{\left(m_{11}+m_{22}\right)^{2}-4}}{2},
\end{align}
or more conveniently
\begin{align}
2\cos\left(KL\right)&=m_{11}+m_{22}=Re\left\{Tr\left[\mathbf{M}\right]\right\}\nonumber\\
&=\cos\left(k_{x}D\right)\cos\left(q_{x}W\right)\nonumber\\
&-\sin\left(k_{x}D\right)\sin\left(q_{x}W\right)\left(\frac{ss'-\sin\phi\sin\theta}{\cos\phi\cos\theta}\right).
\end{align}
This transcendental equation gives the band structure at the
transport problem.

We can also calculate the transmission probability through N
barriers based on the following arguments:
\begin{align}
\left(\begin{array}{c}a_{N}\\b_{N}\end{array}\right)&=\mathbf{M}^{N}\left(\begin{array}{c}a_{1}\\b_{1}\end{array}\right)=\mathbf{T}\left(\begin{array}{c}a_{1}\\b_{1}\end{array}\right),
\end{align}
where
\begin{widetext}
\begin{align}
\mathbf{M}^{N}=\left(\begin{array}{cc}\frac{m_{11}\sin\left(NKL\right)-\sin\left[\left(N-1\right)KL\right]}{\sin\left(KL\right)}&\frac{m_{12}\sin\left(NKL\right)}{\sin\left(KL\right)}\\\frac{m_{21}\sin\left(NKL\right)}{\sin\left(KL\right)}&\frac{m_{22}\sin\left(NKL\right)-\sin\left[\left(N-1\right)KL\right]}{\sin\left(KL\right)}\end{array}\right).
\end{align}
\end{widetext}
In the case that the total flux is conserved, the transmission
probability is expressed as below:
\begin{align}
T&=\frac{T}{T+R}=\frac{1}{1+\frac{R}{T}}=\frac{1}{1+\left|\left(\mathbf{M}^{N}\right)_{21}\right|^{2}}\nonumber\\
&=\frac{1}{1+\left|m_{21}\right|^{2}\frac{\sin^{2}\left(NKL\right)}{\sin^{2}\left(KL\right)}}.
\end{align}

\section{Band structures for magnetic superlattices}
\label{app:band_structure}

As aforementioned, for an infinite number of magnetic barriers,
i.e., magnetic superlattices, the band structures are obtained by
\begin{align}
2\cos\left(KL\right)=Re\left\{Tr\left[\mathbf{M}\right]\right\},
\end{align}
where $K$ is the Bloch wavevector and $L$ is the period of magnetic
superlattices. The band structures of magnetic superlattices
reflects transport properties, especially the nature of the
transporting gap.

\begin{figure}[hbtp!]
\includegraphics[width=8.5cm]{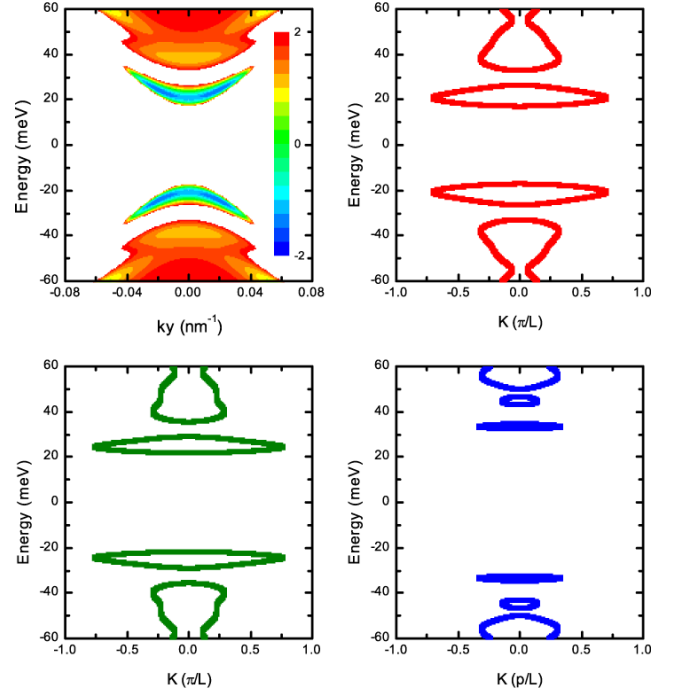}
\caption{ (a) Contour plot of
$\left|Re\left\{Tr\left[\mathbf{M}\right]\right\}\right|\leq2$ as a
function of energy and $k_{y}$. (b)-(d) Energy dispersion versus the
Bloch wavevector for $k_{y}=0$, $2$, and $4$.}\label{fg:dispersion}
\end{figure}

Let us consider a magnetic superlattice characterized by the barrier
width $W=50~nm$, the inter-barrier distance $D=50~nm$, and the
barrier height $\hbar v_{F}eA_{y}/l_{B}=19.8~meV$. In the magnetic
superlattice, the alternating profile of magnetic barriers is taken
into account as shown in Fig. \ref{fg:model}(b). In this case, the
period is given by $L=2(W+D)$. Figure \ref{fg:dispersion} represents
the contour plot of $Re\left\{Tr\left[\mathbf{M}\right]\right\}$ as
a function of Fermi energy and $k_{y}$, and exhibits band structures
as functions of the Bloch wavevector $K$. It is easily found that
the band gap $\sim2\times\hbar v_{F}eA_{y}/l_{B}=39.6~meV$ is
consistent with the transporting gap shown in Fig.
\ref{fg:cond_defect}. Interestingly, there exist other forbidden
gaps in the energy ranges [20:40] or [-40:-20]. Compared to the band
gap between the conduction and valence bands, these gaps depend on
$k_{y}$ values. (see Fig. \ref{fg:dispersion}(b)-(d)) Because of the
$k_{y}$-dependence, the conductance values are incompletely
suppressed down, forming transmission dips around the energy ranges
[20:40] as depicted in Fig. \ref{fg:cond_defect}.

This formalism and the band structures are, of course, valid for
only an infinite array of magnetic barriers. However, if the number
of magnetic barriers are large enough, the qualitative analysis is
expected to be quite consistent with the finite number of barriers.

\end{document}